\documentclass{jkas}

\def\year{2014} 
\def\month{March} 
\def\volume{00} 
\def\issue{0} 
\def\beginpage{1} 
\def\endpage{99} 
\def\received{---} 
\def\accepted{---} 

\date{Received \received ; accepted \accepted}

\title{
    Augmenting \emph{WFIRST} Microlensing With A Ground-Based Telescope Network
}

\author[1,2]{Wei Zhu}
\author[3,1,2]{Andrew Gould}

\affil[1]{Department of Astronomy Ohio State University, 140 W.\ 18th Ave., Columbus, OH 43210, USA; \email{weizhu@astronomy.ohio-state.edu }}
\affil[2]{Max-Planck-Institute for Astronomy, K\"onigstuhl 17, 69117 Heidelberg, Germany}
\affil[3]{Korea Astronomy and Space Science Institute, Daejon 305-348, Republic of Korea;
\email{gould@astronomy.ohio-state.edu }}

\def\corrauthor{
Wei Zhu
}

\def\runningauthor{
Wei Zhu \& Andrew Gould 
}

\def\runningtitle{
    \emph{WFIRST} Plus Ground-based Microlensing Telescope Network
}

\def\keywords{
astrometry -- gravitational microlensing -- planets -- stars: fundamental parameters (mass)
}

\def\abstracttext{
    Augmenting the {\it WFIRST} microlensing campaigns with intensive observations from a ground-based network of wide-field survey telescopes would have several major advantages. 
    First, it would enable full two-dimensional (2-D) vector microlens parallax measurements for a substantial fraction of low-mass lenses as well as planetary and binary events that show caustic crossing features. For a significant fraction of the free-floating planet (FFP) events and all caustic-crossing planetary/binary events, these 2-D parallax measurements directly lead to complete solutions (mass, distance, transverse velocity) of the lens object (or lens system). For even more events, the complementary ground-based observations will yield 1-D parallax measurements. Together with the 1-D parallaxes from \wfirst~alone, they can probe the entire mass range $M\gtrsim M_\oplus$. 
    For luminous lenses, such 1-D parallax measurements can be promoted to complete solutions (mass, distance, transverse velocity) by high-resolution imaging.  This would provide crucial information not only about the hosts of planets and other lenses, but also enable a much more precise Galactic model. 
    Other benefits of such a survey include improved understanding of binaries (particularly with low mass primaries), and sensitivity to distant ice-giant and gas-giant companions of \emph{WFIRST} lenses that cannot be detected by \emph{WFIRST} itself due to its restricted observing windows. Existing ground-based microlensing surveys can be employed if \emph{WFIRST} is pointed at lower-extinction fields than is currently envisaged. This would come at some cost to the event rate.  Therefore the benefits of improved characterization of lenses must be weighed against these costs.
}

\newcommand{\bdv}[1]{\mbox{\boldmath$#1$}}

\def\au{{\rm AU}}

\def\kms{{\rm km}\,{\rm s}^{-1}}
\def\masyr{{\rm mas}\,{\rm yr}^{-1}}
\def\muas{{\mu\rm as}}

\def\mas{{\rm mas}}
\def\lim{{\rm lim}}
\def\sat{{\rm sat}}

\def\max{{\rm max}}
\def\peak{{\rm peak}}
\def\rel{{\rm rel}}

\def\eff{{\rm eff}}

\def\e{{\rm E}}
\def\bpi{{\bdv\pi}}
\def\bmu{{\bdv\mu}}

\def\bLambda{{\bdv\Lambda}}

\def\apj{{ApJ}}
\def\apjl{{ApJL}}

\def\mnras{{MNRAS}}

\def\s{{\rm s}}
\def\base{{\rm base}}
\def\wfirst{\emph{WFIRST}}
\def\eff{{\rm eff}}
\def\w{{\rm W}}
\def\sat{{\rm sat}}
\def\kms{{\rm km~s^{-1}}}
\def\au{{\rm AU}}

\begin{document}
\jkashead 


\section{{Introduction}
\label{sec:intro}}

The proposed {\it WFIRST} mission \citep{Spergel:2015} contains a significant microlensing component, which will plausibly consist of six roughly 72-day continuous campaigns with 15 minute cadence covering about $2.8\,\rm deg^2$ of the Galactic Bulge. The observations will be made from L2 orbit and each campaign will be centered on quadrature, i.e., roughly March 21 and September 21. The observations will be carried out in a broad $H$-band, which is substantially less affected by dust than optical bands. In principle, this permits observations closer to the Galactic plane where the microlensing event rate is almost certainly higher than in the lowest-latitude fields accessible to ground-based $I$-band microlensing surveys.
 
The full power of a space-based survey at L2 can only be realized by complementing it with a deep, high-cadence, near continuous survey, as previous studies have suggested. \citet{gould03}, \citet{han04}, and \citet{yee13} have previously discussed the possibility of microlens parallax measurements by combining observations from Earth and L2 orbit. \citet{han04} demonstrated that combining large format surveys from these two locations would yield microlens parallaxes and so mass measurements for low-mass objects, particularly FFPs. \citet{yee13} focused on the specific challenges of conducting such parallax observations using {\it WFIRST}.

In this work, we provide a quantitative analysis of how the success of \emph{WFIRST} microlensing experiment can be enhanced by complementary ground-based survey observations. The value of the so-called one-dimensional (1-D) parallax measurements is discussed.  These will be substantially more plentiful than the 2-D parallaxes and, more importantly, measurable at all Einstein timescales. We also point out two additional benefits that such ground-based survey observations can provide to \emph{WFIRST}.

In our calculations, we assume these ground-based observations are taken by a survey similar to the Korean Microlensing Telescope Network \citep[KMTNet,][]{Kim:2016,Henderson:2014}. We recognize that adopting such an optical survey comes with a loss in event rate, as it requires that \emph{WFIRST} be pointed at lower-extinction fields than is currently envisaged. Therefore, the benefits of adopting such a survey must be weighed against such a cost. Our methodology can be easily adapted for any specified microlensing surveys, either ongoing or planned.

\section{Microlensing Parallaxes} \label{sec:threeparm}

The standard microlensing light curve normally only yields one single observable that is of physical interest, the event timescale $t_\e$,
\begin{equation} \label{eqn:te}
t_\e \equiv \frac{\theta_\e}{\mu_\rel}\ .
\end{equation}
Here $\mu_\rel=|\bdv{\mu}_\rel|$ is the relative proper motion between the lens and the source, and $\theta_\e$ is the angular Einstein radius
\begin{equation} \label{eqn:thetae}
\theta_\e \equiv \sqrt{\kappa M_{\rm L} \pi_\rel}\ ,
\end{equation}
where $\kappa \equiv 4G/(c^2\au)$ and $\pi_\rel = \au(D_{\rm L}^{-1}-D_{\rm S}^{-1})$ is the lens-source relative parallax. Here $D_{\rm L}$ and $D_{\rm S}$ are the distances to the lens and the source, respectively.

Under certain circumstances, the microlens parallax effect can be measured, which yields the microlens parallax parameter,
\begin{equation} \label{eqn:pievec}
\bpi_\e \equiv \pi_\e \frac{\bmu_\rel}{\mu_\rel};\quad \pi_\e \equiv \frac{\pi_\rel}{\theta_\e}\ .
\end{equation}
This can be done by using a single accelerating observatory \citep{Gould:1992}, or by taking observations simultaneously from two separated sites \citep{Refsdal:1966,Gould:1994b,holz96,gould97}. In the latter case, the microlens parallax vector is given by
\begin{equation} \label{eqn:piesat}
\bpi_\e = \frac{\au}{D_{\sat,\perp}} \left(\frac{\Delta t_0}{t_\e},\Delta u_0\right)\ ,
\end{equation}
where $D_{\sat,\perp}$ is the projected separation between the two observatories evaluated at the peak of the event, and $\Delta t_0=t_{0,\sat}-t_{0,\oplus}$ and $\Delta u_0 = u_{0,\sat}-u_{0,\oplus}$ are the differences in the peak times and impact parameters as seen from the two observatories (we have assumed Earth and one satellite), respectively.

In many cases, however, only the component of the vector $\bpi_\e$ parallel to the opposite direction of acceleration (i.e., away from the Sun for Earth and \emph{WFIRST}) projected onto the sky can be measured with reasonable precision. This component is denoted as $\pi_{\e,\parallel}$
\begin{equation} \label{eqn:piepar}
\pi_{\e,\parallel} \equiv \bpi_\e \cdot \hat{\bf n}\equiv \pi_\e \cos\phi_\pi\ ,
\end{equation}
where $\hat{\bf n}$ is the opposite direction of acceleration projected on the sky, and $\phi_\pi$ is the angle between $\bpi_\e$ and $\hat{\bf n}$.

The parameters $\pi_{\e,\parallel}$ and $\bpi_\e$ are often called 1-D and 2-D parallaxes, respectively. 

To facilitate later discussions, we introduce the vector microlensing parameter $\bdv{\Lambda}$ \citep{dong07}
\begin{equation} \label{eqn:lambda} 
\bdv{\Lambda} \equiv \frac{t_\e \bpi_\e}{\au} = {\pi_\rel/\mu_\rel\over \au}\,
{\bmu_\rel\over \mu_\rel},
\end{equation}
whose amplitude $\Lambda=1/\tilde v$ is the reciprocal of the 
projected transverse velocity.  Because $\bLambda$ is a purely
kinematic quantity, it can 
be used to distinguish between disk events (disk lens + Bulge source) and Bulge events (Bulge lens + Bulge source) \citep{Gould+:1994}.  That is, its (inverse) amplitude is typically
\begin{equation} \label{eqn:diskbulge}
    \tilde v = \left\{ \begin{array}{ll}
            280\kms & {,\ \rm Disk~events} \cr
            1000\kms & {,\ \rm Bulge~events}
    \end{array} \right.
\end{equation}
Here we have adopted $\pi_\rel=0.12~\mas$ and $\mu_\rel=7~\masyr$ as typical values for disk events, and $\pi_\rel=0.02~\mas$ and $\mu_\rel=4~\masyr$ for Bulge events.

In the satellite parallax method, $\bLambda$ is more directly measured than $\bpi_\e$ (Equation~(\ref{eqn:piesat}))
\begin{equation} \label{eqn:lambdasat}
\bdv{\Lambda} = \frac{1}{D_{\sat,\perp}} \left(\Delta t_0,t_\e\Delta u_0\right)\ ,
\end{equation}
because $\Delta t_0$ 
is usually better measured than $t_\e$.  Moreover, as we will show below,
for the special case of L2 space parallaxes (as well as terrestrial parallaxes,
\citealt{gouldyee13}) $t_\e\Delta u_0\rightarrow \Delta t_\eff$ where
$t_\eff \equiv u_0 t_\e$, and $t_\eff$ can be much better measured than $t_\e$.
This is especially true for short timescale events, as $t_\e$ and $u_0$ can be severely degenerate with each other as well as other parameters such as the source flux $F_{\rm s}$.

\section{{\emph{WFIRST}+Ground Parallaxes}
\label{sec:wfirst+groundpie}}

Whenever there are microlens parallax measurements from comparing
the light curves of two observatories, it is also possible to
obtain complementary parallax information from the accelerated
motion of one or both observatories separately.  In the present case,
\emph{WFIRST} orbital parallax will play a major
complementary role to the two-observatory parallaxes that
are made possible by a ground-based observatory (or network of
ground-based observatories).  However, for clarity, we begin by
analyzing the parallax information that can be derived by comparing
the two light curves.

\emph{WFIRST}-Earth microlensing has some features that differ substantially from those two-observatory experiments that have been carried out previously or that are being carried out.  First, since \emph{WFIRST} is a dedicated space-based photometry experiment, it will almost always have essentially perfect measurements relative to the ground. Therefore, the errors in the parallax measurements are very well approximated as those due to the ground observations. Second, for similar reasons, \emph{WFIRST}-selected events will be quite faint as seen from Earth, and therefore the Earth-based photometry errors can be treated as ``below sky'', i.e., independent of flux.  Third, since \emph{WFIRST} will be at L2, its projected motion relative to Earth will be extremely slow, substantially less than $1\,\kms$. This can be compared with typical lens-source projected velocities $\tilde v\sim {\cal O}(100\,\kms)$.  This means that the Einstein timescales $t_\e$ are essentially identical as seen from the two locations.  In particular, it implies that the quantity entering $\Lambda_\perp=t_\e\Delta u_0/D_{\rm sat,\perp}$ can be simplified by
\begin{equation}
t_\e\Delta u_0 = \Delta(u_0 t_\e)-u_0\Delta t_\e \rightarrow \Delta t_\eff
\label{eqn:deltateff}
\end{equation}
where $t_\eff\equiv u_0 t_\e$.  That is, Equation~(\ref{eqn:lambdasat}) becomes
\begin{equation} \label{eqn:lambdasat2}
\bdv{\Lambda} \rightarrow \frac{1}{D_{\sat,\perp}} 
\left(\Delta t_0,\Delta t_\eff\right)\ ,
\end{equation}
as we anticipated above.
In addition, in most
cases, $t_\e$ will in fact be measured from \emph{WFIRST} even if
it cannot be measured from Earth, so that we can then convert
$\bpi_\e = (\au/t_\e)\bLambda$.  From our standpoint, we will therefore
regard measurement of $\bLambda$ as the goal, with the understanding
that this itself will very often yield $\bpi_\e$.  And even when it cannot,
$\bLambda$ is the crucial parameter for distinguishing populations
in any case because it is a purely kinematic variable.
Finally, since \emph{WFIRST} will be at L2,
\emph{WFIRST}-Earth parallaxes are exceptionally
sensitive to short events, which is traditionally the most difficult
regime, i.e., the regime of events generated by very low-mass lenses.  
That is, such events do
not last long enough to make orbital parallax measurements, and their
Einstein radii are too small to permit simultaneous observation
from observatories separated by $\sim \au$, like {\it Spitzer}
and {\it Kepler}.

We analyze the \wfirst-ground parallaxes using Fisher matrices. The full point-lens equation is described by four parameters that are of physical interest, $a_i=(t_0,u_0,t_\e,F_\s)$, and one nuisance parameter $F_\base$
\begin{equation} \label{eqn:plps}
    F(t) = F_\s (A[u(t)]-1) + F_\base\ ,
\end{equation}
where
\begin{equation} \label{eqn:amp}
A(u) = \frac{u^2+2}{u\sqrt{u^2+4}}\ ;\quad u^2 = \tau^2+u_0^2\ ;\quad \tau \equiv \frac{t-t_0}{t_\e}\ .
\end{equation}
The nuisance parameter $F_\base$ is essentially uncorrelated with other parameters, so we ignore it in the following analysis. Under the assumption of uniform observations at a cadence $\Gamma$, the Fisher matrix (i.e., inverse of the covariance matrix) is then given by
\begin{equation} \label{eqn:bij_general}
    b_{ij} = \frac{\Gamma}{\sigma_0^2} \int_{-\infty}^{+\infty} dt \frac{\partial F(t)}{\partial a_i} \frac{\partial F(t)}{\partial a_j}\ ,
\end{equation}
where we have assumed that the observations are below sky so that the
flux error $\sigma_0$ is independent of magnification.
Here
\begin{equation} \label{eqn:partials_general}
\frac{\partial F}{\partial a_i} = \left( \begin{array}{c}
        -F_\s A'\tau/(ut_\e) \cr
        F_\s A' u_0/u \cr
        -F_\s A' \tau^2/(ut_\e) \cr
        A-1 \cr
\end{array} \right)\ ,
\end{equation}
and
\begin{equation} \label{eqn:dadu_general}
A' \equiv \frac{dA}{du} = \frac{-8}{u^2(u^2+4)^{3/2}}\ .
\end{equation}

Although the Fisher matrix cannot be expressed in closed form for the general case, it can be in the high magnification regime, where $A(u)=1/u$ and $A'(u)=-1/u^2$. Below we derive these closed-form expressions in this regime, and provide the analysis of the general case in Appendix~\ref{sec:app1}.

In the high-magnification limit, it can be seen that
\begin{equation} \label{eqn:degen}
    u_0 \frac{\partial F}{\partial u_0} - t_\e\frac{\partial F}{\partial t_\e} + F_\s\left(\frac{\partial F}{\partial F_\s}+1\right) = 0\ .
\end{equation}
That is, in this limiting regime, the parameters $(u_0,t_\e,F_\s)$ are degenerate. Hence, the only way to distinguish them is from the wings of the light curve. This can be a serious problem
for ground-based observations of \emph{WFIRST} targets, since they may be extremely faint and noisy near baseline.

\begin{figure}
\centering
\includegraphics[width=80mm]{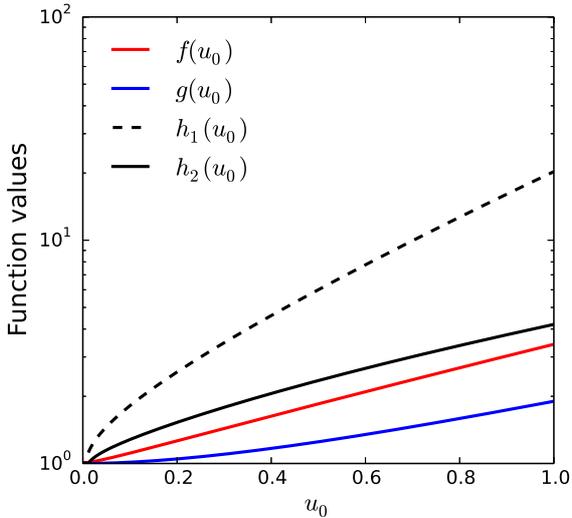}
\caption{Functional forms of $f(u_0)$ (defined as Equation~(\ref{eqn:fu0})), $g(u_0)$ (defined as Equation~(\ref{eqn:gu0})), $h_1(u_0)$ (defined as Equation~(\ref{eqn:h1u0})), and $u_2(u_0)$ (defined as Equation~(\ref{eqn:h2u0})) in the range $0<u_0<1$.
\label{fig:func_u0}}
\end{figure}

However, as stated above we are not actually interested in directly measuring $t_\e$ from the ground. We therefore rewrite Equations~(\ref{eqn:plps}) and (\ref{eqn:amp}) in the high-magnification limit, which has only three parameters \citep{gould96} $a_i=(t_0,t_\eff,F_\peak)$
\begin{equation} \label{eqn:plpslimit}
F(t) = F_\peak Q(t)\ ;\qquad Q(t) = \biggl({(t-t_0)^2\over t_\eff^2} + 1\biggr)^{-1/2}\ .
\end{equation}
Then 
\begin{equation}
{\partial F\over \partial a_i}\rightarrow\left(\matrix{
F_\peak Q^3\tau_\eff/t_\eff\cr
F_\peak Q^3\tau_\eff^2/t_\eff\cr
Q}\right)\ ,
\label{eqn:dlimit}
\end{equation}
where $\tau_\eff\equiv (t-t_0)/t_\eff$.  We then evaluate the inverse covariance matrix, 
\begin{equation} \label{eqn:bij}
b_{ij} = {\pi\over 8}{\Gamma t_\eff F_\peak^2\over \sigma_0^2}
\left(\matrix{
t_\eff^{-2} & 0 & 0 \cr
0 & 3t_\eff^{-2} & 4t_\eff^{-1}F_\peak^{-1} \cr
0 & 4t_\eff^{-1}F_\peak^{-1} & 8F_\peak^{-2}}
\right)\ ,
\end{equation}
and thus the covariance matrix $c=b^{-1}$
\begin{equation}
c_{ij} = {8\over\pi}{\sigma_0^2\over F_\peak^2\Gamma t_\eff}
\left(\matrix{
t_\eff^2 & 0 & 0 \cr
0 & t_\eff^2 & -t_\eff F_\peak/2 \cr
0 & -t_\eff F_\peak/2 & (3/8)F_\peak^2}
\right)\ .
\label{eqn:cij}
\end{equation}
Thus the uncertainties on $t_0$ and $t_\eff$ are
\begin{equation} \label{eqn:sigt0}
\sigma(t_0) = \sqrt{{8\over\pi}\,{t_\e\over \Gamma}}{\sigma_0\over F_\s} u_0^{3/2}g(u_0)\ ,
\end{equation}
\begin{equation} \label{eqn:sigteff}
\sigma_i(t_\eff) = \sqrt{{8\over\pi}\,{t_\e\over \Gamma}}{\sigma_0\over F_\s} u_0^{3/2}h_i(u_0)\ .
\end{equation}
The correction factors $g(u_0)$ and $h_i(u_0)$ ($i=1,2$) allow us to extend these formulae to the general case, and are derived in Appendix~\ref{sec:app1}. We provide two different forms of $\sigma(t_\eff)$: the first is derived by using purely ground-based information, while the second is derived by assuming perfect knowledge of $t_\e$ from \wfirst. These three functions, $g(u_0)$, $h_1(u_0)$, and $h_2(u_0)$ are illustrated in Figure~\ref{fig:func_u0} for $0\le u_0 \le 1$.

In principle, $t_\e$ cannot be known perfectly from \wfirst~for two reasons. First, \wfirst~observations are not perfect, so the associated $t_\e$ measurement has a statistical uncertainty $\sigma_\w(t_\e)$. Second, even if $t_\e$ from \wfirst~can be constrained extremely precisely, $t_\e$ of the same event as seen from Earth is still uncertain to a limited level $\Delta t_\e$, due to the relative velocity between \wfirst~and Earth. However, as we show in Appendix~\ref{sec:app1}, $t_\e$ for ground-based observations can be treated as ``perfectly'' known as long as the uncertainty in $t_\e$ inferred from \wfirst~is smaller, by a certain specified factor (Equation~(\ref{eqn:condition})), than the uncertainty in $t_\e$ from ground-based observations. We have further proved in Appendices~\ref{sec:app1} and \ref{sec:app2} that this condition is almost always satisfied for both the \wfirst~statistical uncertainty $\sigma_\w(t_\e)$ and the \wfirst-Earth systematic offset $\Delta t_\e$.

Therefore, the assumption that $t_\e$ is perfectly known from \wfirst~almost always holds, so that $\sigma_2(t_\eff)$ is mostly what we can get for the uncertainty in $t_\eff$ from ground-based observations.

Equations~(\ref{eqn:sigt0}) and (\ref{eqn:sigteff}) have a number of important implications. First, the two terms entering $\bLambda$ have exactly the same errors in the high magnification limit, namely $\sqrt{8 t_\eff/\pi\Gamma}\sigma_0/F_\peak$. Second, the errors scale strongly with magnification, $\propto u_0^{3/2}$ times the correction factor. This implies a strong magnification bias, so that the much more numerous faint potential sources can relatively easily enter the sample at high magnification. The magnification bias is stronger for $\sigma(t_\eff)$ than for $\sigma(t_0)$, suggesting that $\pi_{\e,\parallel}$ is always better determined than $\pi_{\e,\perp}$. However, comparison of $g(u_0)$ and $h_2(u_0)$ shows that this superiority is relatively modest.

Third, $t_0$ is not correlated with other parameters, and in particular it is not correlated with $t_\eff$. This is simply due to the fact that $\partial F/\partial t_0$ is an odd function of $t$, while the other derivatives are even in $t$. This is true for both Equations~(\ref{eqn:partials_general}) and (\ref{eqn:dlimit}).

Fourth, $t_\eff$ is correlated with other parameters. As the first indication of why this is important, we note that even in the high-magnification limit, $t_\eff$ remains significantly correlated with $F_\peak$ (correlation coefficient $-\sqrt{2/3}$). Hence, for example, if there were independent information about the source flux, the error in $t_\eff$ could be reduced by a factor up to $\sqrt{3}$.

To make a quantitative estimate of the microlens parallax errors, we adopt parameters typical of KMTNet. We assume $\Gamma_\oplus=240\,{\rm day}^{-1}$, i.e., one observation per 2 minutes, for four hours per night (which is the time the Bulge is visible at the midpoint of the \emph{WFIRST} campaigns) at each of three observatories, and 33\% bad weather. We normalize the errors to 0.05 magnitudes at $I=18$.
\footnote{The Vega magnitude system is used in the present work.}
We then find
\begin{equation} \label{eqn:sigmaeval}
    \left[\begin{array}{c}
        \sigma(\pi_{\e,\parallel})\cr
    \sigma(\pi_{\e,\perp})
\end{array}\right]
= \frac{0.52}{\cos\psi}\,\left(\frac{t_\e}{\rm day}\right)^{-1/2}
10^{\frac{I_s-18}{2.5}} u_0^{3/2} 
\left[\begin{array}{c}
    g(u_0) \cr
    h_2(u_0)
\end{array}\right]\ .
\end{equation}
where $\psi$ is the phase of the \emph{WFIRST} orbit relative to quadrature at the peak of the event. We note that because the observations are centered at quadrature, $0.82\leq\cos\psi\leq 1$. At first sight this pre-factor does not look especially promising, particularly given the fact that typical \emph{WFIRST} microlensing sources will be substantially fainter than $I_s=18$.  However, there are three points to keep in mind.  First, we expect $t_\e\sim 1\,$day events to correspond to $M\sim M_{\rm Jup}$ lenses, whose parallaxes would be $\pi_\e\sim 4$ if they lay in the disk and $\pi_\e\sim 1.5$ in the bulge.  Second, 10\% of a ``fair sample'' of events will have $u_0<0.1$ and so errors that are $\gtrsim 30$ times smaller. Third, the sample of events will not be ``fair'', but rather heavily biased toward fainter sources at high-magnification.

Regarding the first point, in order to make clear the measurability of parallax, it is better to express Equation~(\ref{eqn:sigmaeval}) in terms of $\bLambda$, since this is a purely kinematic variable that does not vary with the lens mass
\begin{equation} \label{eqn:sigmaeval2}
    \left[\begin{array}{c}
        \sigma(\Lambda_{\parallel}) \cr \sigma(\Lambda_{\perp})
\end{array}\right]
= \frac{0.30\sec\psi}{1000\,\kms}\, 
\left(\frac{t_\e}{\rm day}\right)^{1/2} 10^{\frac{I_s-18}{2.5}} u_0^{3/2} 
\left[\begin{array}{c}
    g(u_0) \cr h_2(u_0)
\end{array}\right]\ .
\end{equation}
This shows that \emph{WFIRST}-Earth parallaxes become {\it more} sensitive at shorter timescales (at fixed $\Lambda$ or projected velocity $\tilde v$).

\begin{figure}
\centering
\includegraphics[width=90mm]{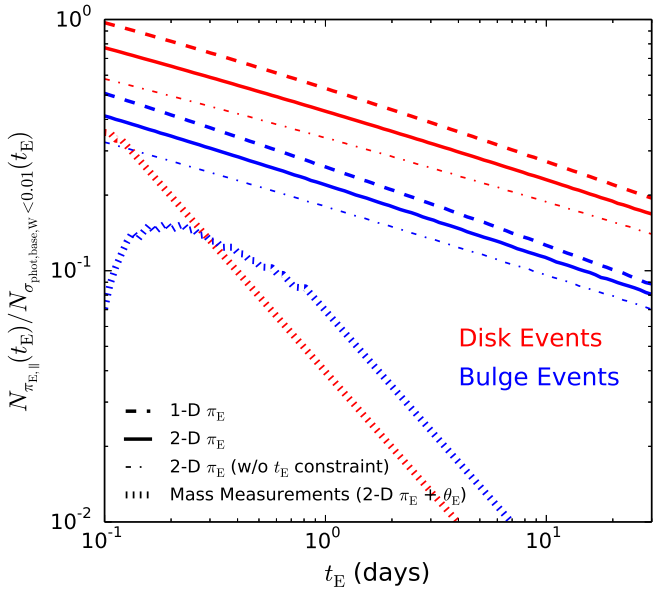}
\caption{The normalized numbers of events with 1-D parallax ($\pi_{\e,\parallel}$) and 2-D parallax ($\bdv{\pi}_\e$) measured better than 5-$\sigma$ for two sets of typical events, respectively. For ``Disk events'', we adopt $\pi_\rel=0.12$ mas and $\mu=7~\masyr$, and for ``Bulge events'', we adopt $\pi_\rel=0.02$ mas and $\mu=4~\masyr$. For each set of typical events, we show two curves for the 2-D parallax measurements, one with perfect $t_\e$ information from \wfirst~(solid lines) and the other without any external $t_\e$ information (dash-dotted lines). The former is more realistic (see Section~\ref{sec:wfirst+groundpie}).
We also show the number of events for which there are both $\bpi_\e$ measurements and $\theta_\e$ measurements (from finite source effects), as discussed in Section~\ref{sec:thetaeval}.
The normalization is the number of events with \wfirst~baseline photometric precision better than 1\%, which corresponds to a baseline magnitude of $H=20.2$ \citep{Gould:2014b}. We assume the \wfirst~microlensing sources follow the luminosity function of Bulge stars derived by \citet{Holtzman:1998}.
\label{fig:fullpar}}
\end{figure}

We illustrate this sensitivity in Figure~\ref{fig:fullpar} by showing the number of events that satisfy $\pi_\e/\sigma(\pi_{\e,\parallel})> 5$ (or $\pi_\e/\sigma(\pi_{\e,\perp})> 5$) for typical Bulge lenses ($\tilde v = \au\mu_\rel/\pi_\rel=1000\,\kms$) and typical disk lenses ($\tilde v = 280\,\kms$). Here we have assumed $A_I=1.5$ and then integrated over the \citet{Holtzman:1998} luminosity function. The Figure is normalized to the number of events with \wfirst~baseline photometric precision better than 1\%. This number can in turn be estimated for any specific \emph{WFIRST} strategy that is either considered or adopted, and of course can be empirically determined from the experiment itself.
 
There are several important points regarding Figure~\ref{fig:fullpar}. First, by incorporating the $t_\e$ information from \wfirst~observations, we are able to increase the number of 2-D parallax detections by a factor $\sim$1.2. Second, the full parallax ($\bdv{\pi}_\e$) curves with and without incorporating $t_\e$ information from \wfirst~lie only about a factor of 1.2 and 1.5 below the $\pi_{\e,\parallel}$ curves, respectively, despite the more serious deteriorations of $h_i(u_0)$ relative to $g(u_0)$ shown in Figure~\ref{fig:func_u0}.
This is because first the measurements are dominated by events with relatively low $u_0\lesssim 0.2$ for which the average ratio $\langle h_1(u_0)/g(u_0)\rangle<2$, and second, for a given source star, this can be compensated by going lower in $u_0$ by a factor $2^{2/3} = 1.6$.

Figure~\ref{fig:fullpar} also shows that the 2-D parallax measurements will be available for a substantial fraction of Jupiter-mass FFPs ($t_\e\lesssim 1\,$day) in the disk and for Earth-mass FFPs ($t_\e\lesssim  0.05\,$day) in the Bulge. Shorter events are in general preferred, but the \wfirst-Earth parallax method implicitly sets a lower limit on the event timescale $t_\e$ that it can probe. This is because the same event must be observable from both \wfirst and Earth, so that $\au/\pi_\e \gtrsim D_{\sat,\perp}$, or
\begin{equation} \label{eqn:telow}
    t_\e \gtrsim 0.02~{\rm day}~\left(\frac{\tilde{v}}{10^3~\kms}\right)^{-1}\cos\psi\ .
\end{equation}
Therefore, Equations~(\ref{eqn:sigmaeval}) and (\ref{eqn:sigmaeval2}) are only valid for disk events with $t_\e\gtrsim0.06$~days and Bulge events with $t_\e\gtrsim0.02$~days.

\section{{One-Dimensional Parallaxes}
\label{sec:1d}}

We now focus attention on 1-D parallaxes. As just mentioned, these can be measured about 1.2 to 1.5 times more frequently than 2-D parallaxes by comparing \wfirst~and ground-based lightcurves. However, our primary reason for this focus is that \emph{WFIRST} can, by itself measure 1-D parallaxes for sufficiently long events. That is, \emph{WFIRST}-Earth and \emph{WFIRST}-only measurements are complementary, being respectively most sensitive in the short $t_\e$ and long $t_\e$ regimes.

Of course, the main disadvantage of 1-D parallaxes is that they appear to be of little practical value. We will show, however, that this assessment is far too pessimistic.

\subsection{{\emph{WFIRST}-only 1-D Parallaxes}
\label{sec:wfirstpiepar}}

We begin by making an estimate of the \emph{WFIRST}-only 1-D parallax errors via Fisher matrix. Because \emph{WFIRST} is observing near quadrature, it is accelerating transverse to the line of sight at 
\begin{equation}
    a_\perp \simeq\au\Omega_\oplus^2\cos\psi;
\qquad
\psi\equiv \Omega_\oplus(t-t_{\rm quad}).
\label{eqn:orbpsi}
\end{equation}
Here $\Omega_\oplus \equiv 2\pi/$yr and $t_{\rm quad}$ is the epoch when the field is at quadrature. In the approximation that the acceleration is constant, this induces a quadratic deviation in the lightcurve, which to lowest order implies a normalized lens-source separation $u(t)$,
\begin{equation}
[u(t)] ^2= \biggl[{t-t_0\over t_\e} + {1\over 2}\pi_{\e,\parallel}
(\Omega_\oplus(t-t_0))^2\cos\psi\biggr]^2 + u_0^2\ .
\label{eqn:uoft}
\end{equation}
This leads to an asymmetric distortion in the magnification \citep{Gould+:1994}.
With this as well as Equation~(\ref{eqn:plps}), one finds that
\begin{equation} \label{eqn:fpiepar}
    \frac{\partial F}{\partial \pi_{\e,\parallel}} = \frac{F_\s \Omega_\oplus^2 t_\e^2 \cos\psi}{2} \frac{A'\tau^3}{u} 
\qquad (\pi_{\e,\parallel}\ll 1)\ .
\end{equation}

Because Equation~(\ref{eqn:fpiepar}) is odd in $t$, the only other microlensing parameter that it couples to is $t_0$. Thus, the Fisher matrix is two-dimensional. To evaluate this, we first specify that \emph{WFIRST} observations will generally be above sky, so that the flux errors scale $\sigma = \sigma_{0,\w}A^{1/2}$, where $\sigma_{0,\w}$ is the error at baseline. As in Equation~(\ref{eqn:bij_general}), we approximate the observations as being at a uniform rate $\Gamma_\w$, and find
\begin{eqnarray}
    b_{ij} &=& \frac{\Gamma_\w t_\e}{\sigma_{0,\w}^2} \int \frac{\partial F}{\partial a_i}\frac{\partial F}{\partial a_j} \frac{d\tau}{A} \cr
    &=& \frac{\Gamma_\w F_\s^2}{4\sigma_{0,\w}^2 t_\e} 
    \left(\begin{array}{cc}
            4G_0 & -2\eta G_1 \cr
            -2\eta G_1 & \eta^2 G_2
    \end{array}\right)\ ,
\end{eqnarray}
where $\eta \equiv \Omega_\oplus^2 t_\e^3 \cos\psi$, and
\begin{equation}
    \left\{ \begin{array}{rcl}
            G_0 & \equiv & \int A'^2 A^{-1} u^{-2} \tau^2 d\tau \cr
            G_1 & \equiv & \int A'^2 A^{-1} u^{-2} \tau^4 d\tau \cr
            G_2 & \equiv & \int A'^2 A^{-1} u^{-2} \tau^6 d\tau
    \end{array}\right.\ .
\end{equation}
Then the covariance matrix is
\begin{equation}
    c_{ij} = b_{ij}^{-1} = \frac{\sigma_{0,\w}^2t_\e}{\Gamma_\w F_\s^2 \eta^2} \frac{1}{G_0G_2-G_1^2} \left(\begin{array}{cc}
            \eta^2 G_2 & 2\eta G_1 \cr
            2\eta G_1 & 4 G_0
    \end{array}\right)\ .
\end{equation}
The uncertainty in $\pi_{\e,\parallel}$ is given by $\sigma(\pi_{\e,\parallel})=c_{22}^{1/2}$. It can be expressed analytically only in the high magnification limit. Therefore, similarly to the case of \wfirst+ground 2-D parallaxes, we introduce a correction factor $f(u_0)$ that approaches unity as $u_0$ approaches zero, and rewrite $\sigma(\pi_{\e,\parallel})$ as
\begin{equation}
    \sigma(\pi_{\e,\parallel}) = 1.77 \frac{\sigma_{0,\w}}{F_\s} \frac{\sec\psi}{\Omega_\oplus^2 t_\e^{5/2} \Gamma_\w^{1/2}} f(u_0)\ ,
\end{equation}
where $1.77=[14/3-32^{1/2}\ln(1+\sqrt{2})]^{-1/2}$ is the result of an analytic calculation, and
\begin{equation} \label{eqn:fu0}
f(u_0) \equiv 1.13\sqrt{\frac{G_0}{G_0G_2-G_1^2}}\ .
\end{equation}
This function is also illustrated in Figure~\ref{fig:func_u0}. The deterioration toward higher $u_0$ is primarily due to the fact that the high-mag peak contributes the majority of the information about $\pi_{\e,\parallel}$.  Secondarily, $t_0$ becomes increasingly correlated with $\pi_{\e,\parallel}$ at higher $u_0$.

Adopting a cadence of $\Gamma_\w=100~{\rm day}^{-1}$ and assuming 0.01 mag errors at $H=20.2$ \citep{Gould:2014b}, this yields
\begin{equation}
    \sigma(\pi_{\e,\parallel}) = \frac{0.017}{\cos\psi}10^{\frac{H_\s-20}{5}} \left({t_\e\over 10\,\rm day}\right)^{-5/2}f(u_0)\ .
\label{eqn:sigpieparw3}
\end{equation}
For an intuitive understanding of the relevance of this error bar to the parallax measurement, it is best to express it in terms of $\Lambda_\parallel$
\begin{equation}
    \sigma(\Lambda_\parallel) = {0.10\,\sec\psi\over 1000\,\kms}10^{\frac{H_s-20}{5}}
\biggl({t_\e\over 10\,\rm day}\biggr)^{-3/2}f(u_0)\ .
\label{eqn:sigpieparw4}
\end{equation}

From Equations~(\ref{eqn:sigpieparw3}) and (\ref{eqn:sigpieparw4}), it is clear that \emph{WFIRST} will make very good $\pi_{\e,\parallel}$ measurements for long events but will do much worse for short events. For example, for $t_\e=2$~days, the pre-factor in Equation~(\ref{eqn:sigpieparw4}), goes from 0.10 to 1.1.

\subsection{{Combined Orbital and Two-Observatory 1-D Parallaxes}
\label{sec:combinedpiepar}}

Comparison of Equations~(\ref{eqn:sigmaeval2}) and (\ref{eqn:sigpieparw4})
shows that the two approaches to obtaining 1-D parallaxes are complementary,
with precisions $\sigma(\Lambda_\parallel)\propto t_\e^{1/2}$ for 
\emph{WFIRST}-Earth parallaxes and $\sigma(\Lambda_\parallel)\propto t_\e^{-3/2}$ 
for \emph{WFIRST}-only parallaxes.  The normalizations 
($I_S=18$ in the first and $H_s=20$ in the second) may appear deceptive,
particularly because typical sources will have $(I-H)\sim 0.8 + E(I-H)\sim 1.8$
assuming $E(I-H)\sim 1$.  However, one must bear in mind that that
the two-observatory formula has a very strong dependence on $u_0$, whereas
the \emph{WFIRST}-only formula is, by comparison, almost flat in $u_0$.

To investigate this further, we consider the combined impact of both measurements, assuming $A_H=0.5$ and (as before) a \citet{Holtzman:1998} luminosity function and $A_I=1.5$. We denote the precision derived by combining these two formulae by $\sigma_{\rm full}(\pi_{\e,\parallel})$.

\begin{figure*}
\centering
\includegraphics[width=80mm]{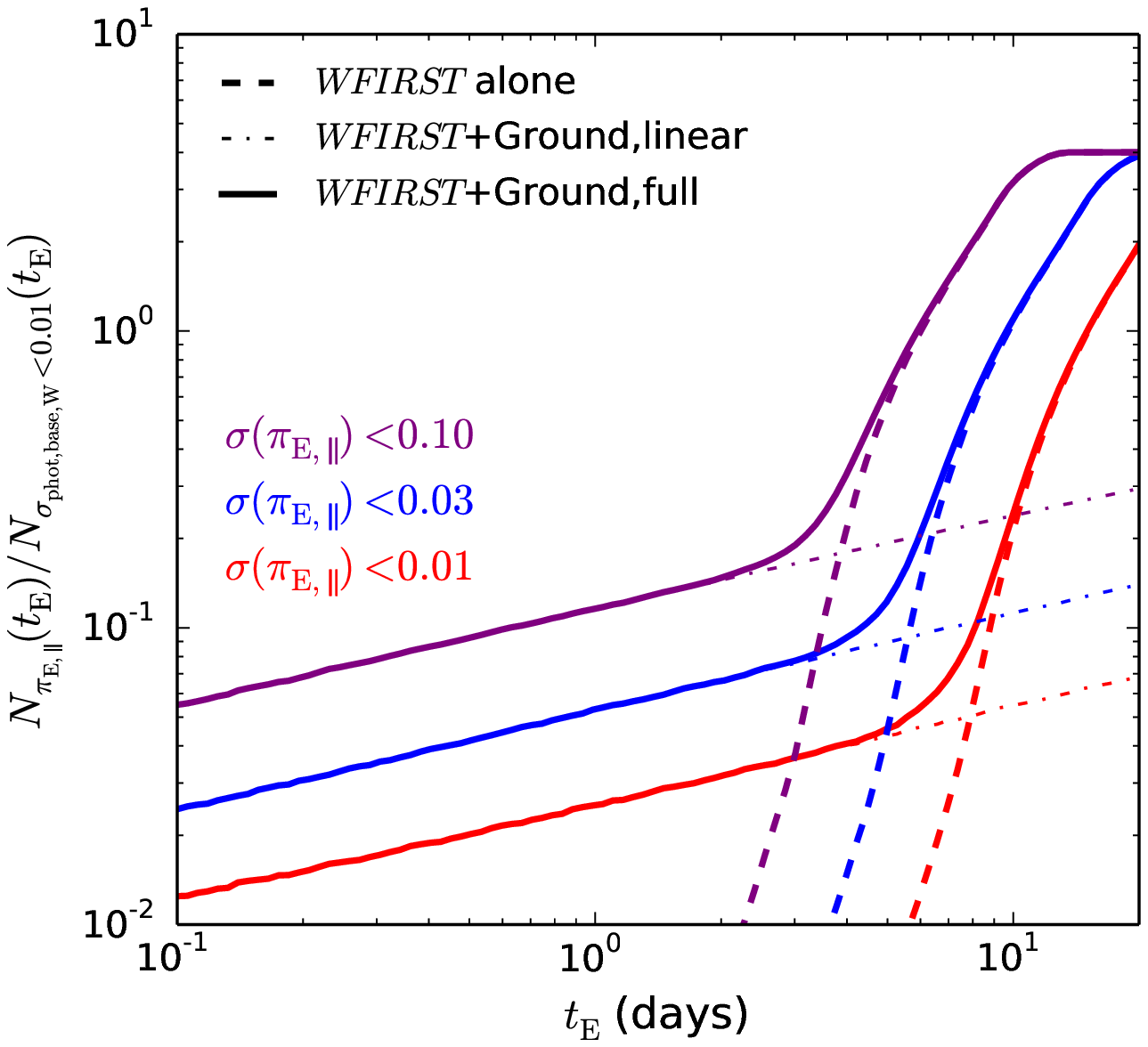}
\includegraphics[width=80mm]{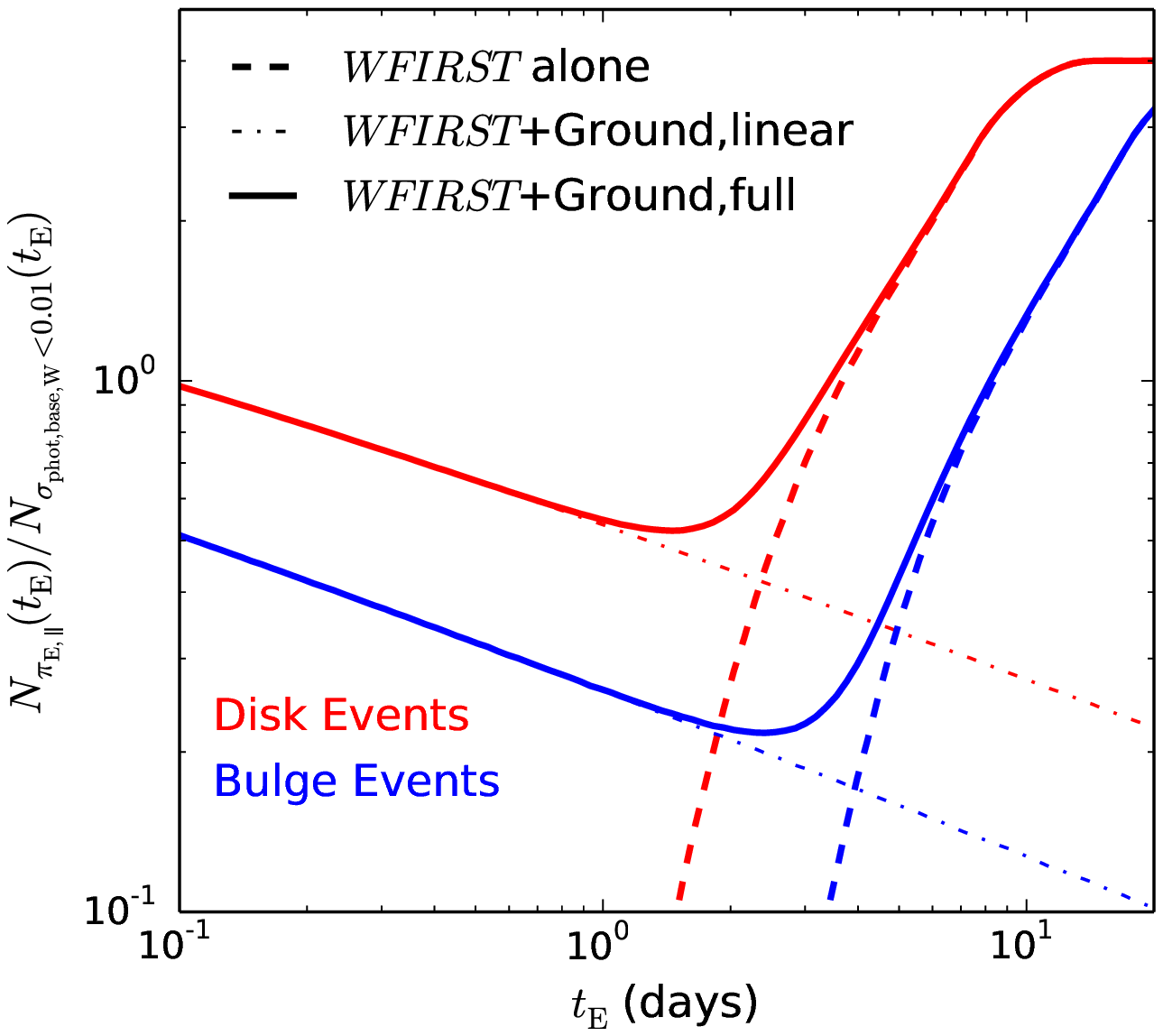}
\caption{The normalized numbers of events with 1-D parallax ($\pi_{\e,\parallel}$) measurements better than given precisions for \emph{WFIRST}-only observations (dashed lines), \emph{WFIRST} plus ground-based survey observations (dash-dotted lines for linear approximation and solid lines for full derivation). The left panel shows curves with absolute measurement uncertainties ($\sigma(\pi_{\e,\parallel})$). The right panel shows curves with 5-$\sigma$ detections ($\pi_{\e,\parallel}/\sigma(\pi_{\e,\parallel})=5$) for typical disk events ($\pi_\rel=0.12$ mas, $\mu=7~\masyr$) and Bulge events ($\pi_\rel=0.02$ mas, $\mu=4~\masyr$). The normalization is the same as that used in Figure~\ref{fig:fullpar}. That is, the number of events with \wfirst~baseline photometric precision seen by \emph{WFIRST} better than 1\%. 
    The flattening for $t_\e>10$~days is caused by our restriction that $u_0\le1$, but it is very likely real for two reasons. First, $\pi_{\e,\parallel}$ measurements will be quite difficult for $u_0>1$, as the correlation between $t_0$ and $\pi_{\e,\parallel}$ becomes significant (correlation coefficient $r(t_0,\pi_{\e,\parallel})\ge0.65$ for $u_0\ge1$); and second, many events with $t_\e>10$ days will not be fully covered by \wfirst~observations.
\label{fig:piepar}}
\end{figure*}

We show in Figure~\ref{fig:piepar} the numbers of events with 1-D parallax measurements better than a specified (absolute and relative) precision for \emph{WFIRST}-only and \emph{WFIRST} plus ground observations, respectively. The normalizations are again to the number of events with \emph{WFIRST} baseline precisions of 1\%. Figure~\ref{fig:piepar} demonstrates the importance of ground-based observations in measuring 1-D parallaxes: without these observations, \emph{WFIRST} will not be able to make any meaningful 1-D parallax measurements for short timescale events.

\subsection{Xallarap and Lens Orbital Motion} \label{sec:xallarap}

Measurements of $\bpi_\e$ made from a single observatory can be corrupted by xallarap (motion of the source about a companion) and lens orbital motion (motion of the lens about a companion), whereas those derived from comparison of contemporaneous measurements from two observatories cannot.  This is because the basis of single-observatory $\bpi_\e$ measurements is the accelerated motion of the observer, which can in principle be perfectly mimicked by accelerated motion of the source (or the lens). Below we only consider the acceleration of the source (i.e., xallarap effect), but our methodology applies to the other case as well.

In the case of complete 2-D $\bpi_\e$ measurements, xallarap-dominated acceleration effects can in principle be distinguished from the parallax effects from their orbital period and the direction of their implied angular momentum vector.  That is, if the effects of parallax are mistakenly attributed to xallarap, then the xallarap solution will lead to a companion with a 1-year period and orbital axis that is exactly aligned to that of the Earth's (projected on the plane of the sky) \citep{Poindexter:2005}.

This purely internal test fails completely, however, for 1-D parallaxes, unless the period is so short that a 2-D xallarap solution can be reliably extracted from the data.  The contamination of 1-D parallaxes due to xallarap has never previously been estimated, probably due to the limited previous interest in 1-D parallaxes themselves.

The contamination due to xallarap can be rigorously calculated under the assumption that the multiplicity properties (masses and semi-major axes) of companions to the microlensed sources are similar to those of solar-type stars in the solar neighborhood. We first note that the amplitude of Earth's acceleration relative to the projected Einstein radius $\tilde r_\e\equiv \au/\pi_\e$ is (at quadrature) $\tilde A = (GM_\odot/\au^2)/\tilde r_\e$.  The component of this acceleration entering $\pi_{\e,\parallel}$ is $\tilde A\cos\phi_\pi$.

By the same token, the acceleration due to a companion of mass $m$ and semi-major axis $a$ relative to the Einstein radius projected on the source plane $\hat r_\e\equiv D_{\rm S}\theta_\e$, is $\hat A = (Gm/a^2)/\hat r_\e$.  And the component that contributes to the asymmetry of the event is $\hat A\cos\phi_\xi$, where $\phi_\xi$ is the angle between the lens-source relative motion and the instantaneous acceleration of the source about its companion.  Hence, the ratio of the xallarap-to-parallax signals contributing to this asymmetry is
\begin{equation}
{|\xi_\parallel|\over|\pi_{\e,\parallel}|}=
{\hat A|\cos\phi_\xi|\over \tilde A|\cos\phi_\pi|}
= {m\over M_\odot}\,
\biggl({a\over \au}\biggr)^{-2}{D_{\rm L}\over D_{\rm LS}}
{|\cos\phi_\xi|\over |\cos\phi_\pi|},
\end{equation}
where $D_{\rm LS}\equiv D_{\rm S} - D_{\rm L}$ is the distance between the lens and the source.

\begin{figure}
\centering
\includegraphics[width=80mm]{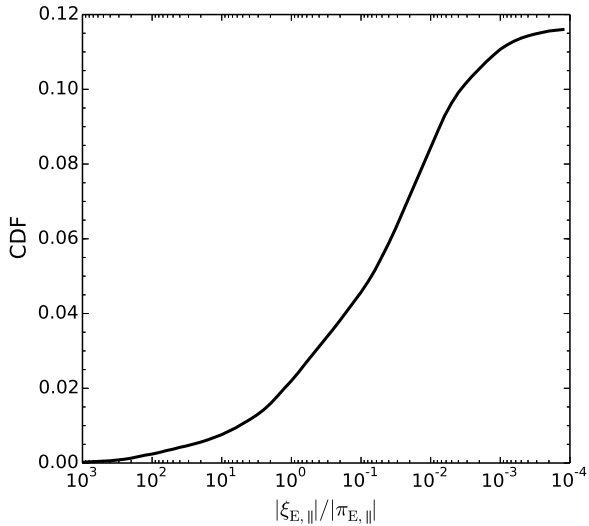}
\caption{The fraction of Bulge events ($D_{\rm L}/D_{\rm LS}=8$) with parallax signal affected by the xallarap effect to a given precision.
\label{fig:xallarap}}
\end{figure}

To evaluate the distribution of this ratio for Bulge lenses, we first adopt $D_{\rm L}/D_{\rm LS}=8$.  We then consider the ensemble of binary companions in Figure 11 of \citet{Raghavan:2010}, which is a complete sample of companions for 454 G-dwarf primaries.  We restrict attention to the 53 companions with semi-major axes $0.2<a/\au<30$ on the grounds that the handful of closer companions would be recognized as such from oscillations in the lightcurve, while those farther away would induce asymmetries that are too small to measure.  We allow $\phi_\pi$ to vary randomly over a circle and $\phi_\xi$ to vary randomly over a sphere.  Figure~\ref{fig:xallarap} shows that xallarap contamination is very serious ($>$100\%) for about 2\% of Bulge lenses and is significant ($>$10\%) for about 4.5\%. In addition, we note that there is a comparable contamination from the acceleration of the lenses due to their companions (both with and without binary lightcurve signatures).

Therefore, with \emph{WFIRST} alone up to 9\% of all events will show false parallax detections. 
Such contamination would be removed by a complementary ground-based survey for relatively short events ($t_\e\lesssim3$ days). For longer events, this contamination cannot be removed completely due to the limited power of \wfirst-Earth 1-D parallax, but the study of this contamination in the short event sample allows to better interpret the 1-D parallax measurements for these longer events.

\subsection{{Value of $\pi_{\e,\parallel}$ Measurements}
\label{sec:pieparval}}

A large sample of 1-D parallaxes with well-understood selection has two major uses.  First, it can be directly analyzed to simultaneously derive a Galactic model and the lens mass function. Second, it allows one to derive the complete solution (namely individual masses, distances and transverse velocities) once combined with the measurement of the lens-source relative proper motion $\bmu_\rel$, which can be obtained for some events with \emph{WFIRST} alone and even more events with follow-up imaging. These two applications are discussed in turn.

In both cases, we show that the addition of short $t_\e$ events from combining ground and space observations will play a critical role in understanding the low-mass end of the mass spectrum, i.e., brown dwarfs (BDs) and FFPs.

{\subsubsection{Statistical Studies Using 1-D Parallax Samples}
\label{sec:statpiepar}}

At present, large microlensing samples with well-understood selection are characterized by a single microlensing variable, the Einstein timescale $t_\e$, i.e., no parallax information. Although not usually thought of this way, we dub this case as a ``0-D parallax measurement'' in order to contrast it to 1-D and 2-D parallaxes. That is, in the three cases, the available information consists of $(t_E)$, $(t_E,\pi_{\e,\parallel})$, $(t_E,\pi_{\e,\parallel},\pi_{\e,\perp})$.

The statistical interpretation of a 0-D parallax sample requires a Galactic model to constrain six of the input variables (lens and source distances and transverse velocities) to obtain information about the seventh (lens mass function).  This situation is not as bad as it may first seem.  The three source properties are well understood statistically from direct observations, even if these source properties are not measured for each individual event, and Galactic models are constructed based on a wide variety of very good data. \citet{sumi11} used this technique to infer the existence of a population of FFPs. Nevertheless, uncertainties in Galactic models remain considerable, and hence it would be valuable if microlensing studies could further constrain them rather than propagating them. Moreover, even if the Galactic model were known perfectly, the statistical precision of microlensing studies is greatly reduced by the requirement of deconvolving three parameters to learn about the one of greatest interest.

\citet{hangould95} argued that individual lens mass and distance would be much more tightly constrained by measuring the full parallax vector $\bpi_\e$ in addition to $t_\e$, and \citet{21event} showed that this was in fact the case for their sample of microlenses with \emph{Spitzer} parallaxes.  Their work still incorporated Galactic models, but the mass/distance constraints were dramatically improved compared the case with timescales alone. See also \citet{gould00}.

The 0-D parallaxes of \citet{sumi11} and the 2-D parallaxes of \citet{21event} have had different applications. Specifically, the great value of the first is that it could make statistical statements about low-mass objects, while that of the second was the greatly improved precision of individual lenses. In particular, the 2-D study could not make any statement about low-mass objects because the \emph{Spitzer} sample was strongly biased against short events.

For \emph{WFIRST} 1-D parallaxes, the situation is clearly intermediate between the 0-D parallaxes of \citet{sumi11} and the 2-D parallaxes of \citet{21event}. 
The great potential value of \emph{WFIRST} 1-D parallaxes (i.e., much stronger statistical statement about BDs and FFPs) would be almost completely lost if these substellar objects were systematically excluded from the 1-D parallax sample. Given the $\sigma(\Lambda_\parallel)\propto t_\e^{-3/2}$ behavior of 
Equation~(\ref{eqn:sigpieparw4}) 
(also see Figure~\ref{fig:piepar}), this is exactly what would happen in the absence of a ground-based complementary survey.  The addition of such a survey would enable 1-D parallax measurements across the entire range of timescale $t_\e\gtrsim 2\,$hrs for disk lenses.

\subsubsection{Complete Solutions From $\pi_{\e,\parallel}$ Plus $\bmu_\rel$} \label{sec:complete}

A complete solution of the lens (or the lens system) can be derived by combining the $\pi_{\e,\parallel}$ measurement from the light curve and the $\bmu_\rel$ measurement from high-resolution imaging \citep{Ghosh:2004,Gould:2014}, based on the definition of $t_\e$ (Equation~(\ref{eqn:te})) and the fact that $\bpi_\e$ and $\bmu_\rel$ are along the same direction. 
\footnote{Because $\pi_{\e,\parallel}$ is measured in the geocentric frame but $\bmu_\rel$ is measured in the heliocentric frame, one must be careful when combining these two measurements. See \citet{Gould:2014} for more details.}

The measurement of $\bmu_\rel$ can be done by \emph{WFIRST} alone, using its $\sim$40,000 high-resolution images, for relatively luminous ($\gtrsim$ early M-dwarfs) lenses \citep{Bennett:2007}. Late-time high-resolution follow-up imaging can extend the domain of coverage to all luminous lenses (i.e., $M> 0.08\,M_\odot$) and can also improve the precision, as well as resolving certain ambiguities that we discuss in Section~\ref{sec:xallarap}.  Such imaging has already achieved important results using {\it HST} \citep{Alcock:2001,Bennett:2015} and Keck \citep{Batista:2015}, but will potentially be much more powerful using next generation telescopes \citep{Gould:2014,Henderson:2015}

Before proceeding, one may ask why one would need microlensing mass and distance measurements for a (subsequently) resolved lens, since photometric estimates would then be available.  As pointed out by \citet{Gould:2014}, the answer is two-fold.  First, one would always prefer measurements to estimates.  Second, high-resolution imaging alone may not identify the correct microlenses. Because roughly 2/3 of all stars are in binaries, and since the lensing cross section scales as $\theta_\e\propto M^{1/2}$, of order 1/3 of the events due to binaries will in fact be generated by the lower-mass (and generally fainter) companion.  In the majority of such cases, the star that actually generated the event will not be visible, because it is either unresolved or dark.  Hence, for tens of percent of cases, the lens would be misidentified by such simple imaging.  \citet{Gould:2014} shows how these cases can be identified and resolved by a combination of 1-D parallaxes and imaging.

A large homogeneous sample of events with both $\pi_{\e,\parallel}$ and $\bmu_\rel$ measurements has several benefits. First, it would permit one to {\it derive} a Galactic model (rather than assuming one).  Second, it would permit one to measure the lens mass function over the range of masses that are probed. Third, for all the planetary events in this sample, one would gain a precise mass measurement of the host and thus (in almost all cases) of the planet.

Even without a complementary ground-based survey, all of these benefits could be derived at least partially from \emph{WFIRST} observations themselves.  However, these observations alone would probe only the upper half or so of the mass function, partly because \emph{WFIRST}-only 1-D parallax measurements require long-timescale events, and partly because \emph{WFIRST} proper motion measurements require luminous lenses. Similarly, the masses of planets orbiting low-mass hosts would remain undetermined.

Hence, complementary ground-based survey observations are crucial to probe the low-mass stellar and substellar lenses.

\subsection{{Value of $\bpi_\e$ Measurements}
\label{sec:pieval}}

As discussed following Equation~(\ref{eqn:sigmaeval2}), the 2-D parallax measurements are biased toward substellar objects.  These are exactly the objects that are inaccessible to the conversion of 1-D parallaxes to complete solutions that was discussed in Section~\ref{sec:complete}. These 2-D parallax measurements enable us to directly derive the complete solutions purely from the microlensing light curves for the roughly 50\% of all planetary and binary events in which $\theta_\e$ can be measured \citep{Zhu:2014}. See Section~\ref{sec:binplan} for details. While most of these 2-D parallax measurements in single-lens events will not yield complete solutions (but see Section~\ref{sec:rho}), \citet{21event} have shown that 2-D parallax measurements, when combined with a Galactic model, give tight constraints on mass and distance, particularly for disk lenses.  This is important not only to further constrain the substellar mass function relative to the analysis that is possible based on 1-D parallaxes (Section~\ref{sec:statpiepar}), but also for a more detailed understanding of individual objects.

\section{{Einstein Radius Measurements: $\theta_\e$}
\label{sec:rho}}

The particular interest of the $\theta_\e$ measurement is that if there is also a measurement of $\bpi_\e$, then it yields a complete solution (i.e., mass, distance and transverse velocity).


\subsection{{\emph{WFIRST}-only Einstein-Radius Measurements}
\label{sec:wfirstthetae}}

In the case of single-lens events, if the lens transits the source, then the light curve is distorted by the finite source effect, which yields $\rho \equiv \theta_*/\theta_\e$, the ratio of the source radius to the angular Einstein radius. Because $\theta_*$ can be determined from the dereddened color and magnitude of the source \citep{ob03262}, this yields a measurement of $\theta_\e$  and so also of the proper motion $\mu_\rel=\theta_\e/t_\e$ \citep{Gould:1994a}.  Such transits occur with probability $\sim\rho$,
\footnote{For ground-based observations, $u_0<\rho$ is required in order to detect the finite-source effect, but for \wfirst, because of its much better photometry and the unblended source (almost always), the maximum allowed $u_0$ can be somewhat bigger than $\rho$.}
which is typically of order $\sim 3\times 10^{-3}$ for main-sequence sources and and $\sim 3\times 10^{-2}$ for giant sources.  

When the observational bias is taken into account, the number of single-lens events with finite-source effects should be more than what one would naively estimate based purely on the above transit probability. Single-lens events with finite-source effects can reach extremely high magnifications, so they can be detected even though the source stars at baseline are extremely faint. For example, an M6 dwarf in the Bulge has $H=26$ (assuming an extinction of $A_H=0.5$) and an angular radius $\theta_\star=0.06\mu$as, and can be magnified in brightness by $A_{\rm max}=2/\rho$ when a lens transits exactly through its center. For a Neptune-mass Bulge lens, this event will reach $H=21$ at its peak. \emph{WFIRST} can therefore obtain on average one observation during the entire transit with a photometric precision of 1.4\%, which would yield a precise measurement of $\rho$.
\footnote{We use \emph{Spitzer} event OGLE-2015-BLG-0763 as an example to prove this. There was only one \emph{Spitzer} observation with photometric precision of 1.2\% when the lens transited the source of OGLE-2015-BLG-0763, but the uncertainty on $\rho$ is already limited to 2.3\% level \citep{Zhu:2015}.}
Hence if the lens is more massive than Neptune, then essentially all luminous stars in the Bulge can serve as source stars for events with measurable finite-source effects.  For Earth-mass Bulge lenses, 2\% precision can be reached for individual measurements of M4 ($M\sim 0.25\,M_\odot$, $R\sim 0.25\,R_\odot$) sources, yielding similar quality $\theta_\e$ determinations (since there are on average 2 measurements per transit). Therefore, although small stars are disfavored by their small angular size, they contribute more significantly to the number of single-lens events with finite-source effects, than they do to the number of all detectable single-lens events.

\subsection{{\emph{WFIRST}+Ground Einstein-Radius Measurements} \label{sec:wfirst+groundthetae}}

In this subsection, we wish to understand how the addition of a ground-based network contributes to the frequency of such measurements. The answer is: quite modestly.  There are two issues.  First, an aggressive ground-based network will only observe the \emph{WFIRST} field about half the time.  Since the source crossings typically last only about 1--2 hours, it is this instantaneous coverage that matters rather than daily coverage.  Second, if the source size, projected on the observer plane ($\rho\au/\pi_\e$) is larger than the Earth-satellite projected separation $D_{\sat,\perp}$, then the probability that the source will transit at least one of the two observers is not substantially increased.  That is, if the ground observatory is taking observations at the time of the peak, it can roughly double this probability provided that
\begin{equation}
{\rho\au\over \pi_\e D_{\sat,\perp}}=100{\theta_*\over \pi_\rel} \sec{\psi} < 1.
\label{eqn:rho}
\end{equation}
That is,
\begin{equation}
    \pi_\rel > 60\,\muas\biggl({\theta_*\over 0.6\,\muas}\biggr)^{-1} \sec{\psi}\ ,
\label{eqn:rho2}
\end{equation}
where we have normalized to the angular radius of a typical solar-type source.  Thus, for main sequence sources, the simultaneous ground-based observations contribute $\theta_\e$ measurements for disk lenses but not Bulge lenses.  The situation is even less favorable for sub-giant and giant sources, which are a small minority of all sources but a larger fraction of all finite-source events due to their larger size \citep{Zhu:2015}.  In brief, we expect that the ground-based survey will add only 10--20\% to the rate of $\theta_\e$ measurements.  This is small enough to ignore for present purposes.

\subsection{{Value of $\theta_\e$ Measurements}
\label{sec:thetaeval}}

The main value of $\theta_\e$ measurements, which come primarily from \emph{WFIRST} observations alone, is derived from combining them with the vector $\bpi_\e$ measurements, which come primarily from combining \emph{WFIRST} and ground-based survey data. We therefore evaluate the conditional probability that $\theta_\e$ is measurable given that $\bpi_\e$ is measured.  

The impact parameter as seen from Earth, $u_{0,\oplus}$, that would allow for a $\bdv{\pi}_\e$ measurement is limited by
\begin{equation}
    (S/N)_{\rm th} \le \frac{\pi_\e}{\sigma(\pi_{\e,\perp})} = \frac{\Lambda D_{\sat,\perp}}{\sigma_2(t_\eff)}\ ,
\end{equation}
where $(S/N)_{\rm th}$ is the threshold for claiming a reliable detection. By approximating $u_0^{3/2}h_2(u_0)\propto u_0^\alpha$ ($\alpha\approx1.7$) in Equation~(\ref{eqn:sigteff}), we can derive the maximum allowed impact parameter $u_{0,\oplus,\max}$
\begin{equation}
    u_{0,\oplus,\max} \propto \left(\frac{\Lambda D_{\sat,\perp}}{(S/N)_{\rm th}}\right)^{1/\alpha} \left(\frac{F_\s}{\sigma_0}\right)^{1/\alpha} \left(\frac{\Gamma_\oplus}{t_\e}\right)^{1/(2\alpha)}\ .
\end{equation}
The impact parameter as seen from \wfirst~is given by $u_{0,\w} = u_{0,\oplus}+\Delta u_0$, where $\Delta u_0$ is related to $\pi_{\e,\perp}$ by Equation~(\ref{eqn:piesat}). For $0.1$~days$\le t_\e \le 3$~days and below sky sources as seen from Earth, one can easily prove that $\Delta u_0 \lesssim u_{0,\oplus,\max}$, so that the maximum allowed impact parameter as seen from \wfirst, $u_{0,\w,\max}$, is similar to $u_{0,\oplus,\max}$. Then the conditional probability can be estimated as
\begin{equation}
    P = \frac{\rho}{u_{0,\w,\max}} = \frac{\theta_\star}{\mu_\rel t_\e u_{0,\w,\max}}\ .
\end{equation}
The stellar angular radius $\theta_\star$ is also related to the stellar flux $F_\s$: $\theta_\star \propto F_\s^\beta$, and we take $\beta\approx0.3$ as a reasonable value for $I$ band. By only keeping $F_\s$ and $t_\e$, we find
\begin{equation}
    P \propto F_\s^{\beta-1/\alpha}~t_\e^{1/(2\alpha)-1} \approx F_\s^{-0.3}~t_\e^{-0.7}\ .
\end{equation}
Hence, this conditional probability is only weakly dependent on source
flux and has a substantially stronger dependence on Einstein timescale.

We adopt $\theta_* = 0.15\,\muas\,10^{0.12(23.5-I_0)}$ and use 
$\sigma_2(t_\eff)\propto u_0^{3/2}h_2(u_0)$ to evaluate $P$ numerically and
show the results in Figure~\ref{fig:fullpar}.  In doing so, we
do not count any events for which $P>1$, since this implies that
the source is too big to permit high-enough magnification for
$t_\eff$ to be measured.    Figure~\ref{fig:fullpar} shows that complete
mass measurements peak at roughly 0.2 days for Bulge lenses and 0.1 days
for Disk lenses, where in both cases they constitute roughly half of the 
2-D parallax measurements.  

\section{{Complete Solutions for Binary and Planetary Events}
\label{sec:binplan}}

Finally, although the complete solutions ($\bpi_\e$ plus $\theta_\e$) 
for isolated stellar-mass lenses will be rare, such measurements are much more
common for planetary and binary events, because roughly half of the
recognizable such events will show steep features due to caustic crossings
and cusp crossings. 
For these, measurement of
$\theta_\e$ by \emph{WFIRST} will be virtually automatic.  Moreover,
these events will also have greatly enhanced 2-D parallax measurements
by three different channels.

First, \citet{angould01} argued that caustic crossing events yield
full 2-D parallaxes much more easily than events without sharp
features (either single-lens events or non-caustic-crossing multiple
lenses).  The sharp features break the continuous degeneracies among
the parameters that are even in $t$.  They also break the
symmetry of the lightcurve, which is the fundamental
cause of the fourth-order time dependence of $\pi_{\e,\perp}$
found by \citet{smp03} for single-lens events.  
While there has been no firm proof that this
is the case, there is substantial circumstantial evidence from the
high fraction of planetary lenses with well-determined parallaxes relative
to single lens events of similar $(u_0,t_\e,f_s)$.  Hence, it is likely
that \emph{WFIRST} will by itself measure many 2-D parallaxes of 
caustic-crossing binary and planetary events, particularly for those
with timescales $t_\e\gtrsim 7\,$days, for which the corresponding
single-lens events show good 1-D parallax measurements. 
See Figure~\ref{fig:piepar}.

Second, events with two caustic crossings observed from both \emph{WFIRST}
and the ground can also yield 2-D parallax determinations based on two
$\Delta t_{\rm cc}$ measurements, one at each crossing.  Here $\Delta t_{\rm cc}$ 
is the difference of the caustic crossing time as seen from \emph{WFIRST}
and the ground.  When they proposed the method of parallax observations
of caustic crossings, \citet{HardyWalker:1995} already recognized
that these yield only 1-D information because  small displacements 
along the direct parallel to the caustic do not generate substantial
changes in the lightcurve.  This is in striking contrast to single
lenses, for which displacements in both directions induce effects
of the same order.  See Figure~\ref{fig:func_u0}. 
\citet{graff02} pointed out that this problem
could be solved if there were two caustic crossings (which is typical
of most events, i.e., entrance and exit) provided that the caustics
themselves are not approximately parallel.  \citet{ob151050} and 
\citet{ob151285} showed in practice that binary events with only one
$\Delta t_{\rm cc}$ measurement are subject to both discrete and
continuous degeneracies.

One shortcoming of this approach with regard to \emph{WFIRST} is that, 
given a network of three southern hemisphere telescopes, 
each caustic crossing can be observed from the ground with approximately
50\% probability simply because the bulge is visible from each
observatory only about 4 hours per night, averaged over the
\emph{WFIRST} campaign.  This emphasizes the importance of having
a network, but also of increasing it to as many independent
locations as possible.  For the example of the KMTNet network, adding 
a node in Hawaii would be valuable.

Yet a third channel would be to combine the 
\emph{WFIRST} 1-D orbital parallax measurement with one $\Delta t_{\rm cc}$
measurement, for the case that there is only one, either because
the second crossing was missed from the ground or because
the geometry of the event only had one crossing (e.g., \citealt{ob151285}).
As in the previous approach there are constraints on the orientation
of the caustic: it cannot be too close to perpendicular to $\hat {\bf n}$.
This method may be the most frequently employed for shorter
events, because, if the probability of catching a single crossing is
$p=50\%$, then the ratio of one measured $\Delta t_{\rm cc}$ to two such
measurements is $(2/p -1)\rightarrow 3$.

A full analysis of the measurability of such parallaxes lies well
beyond the scope for the present work but we believe that this
should be actively investigated.

\section{{Auxiliary Benefits of \emph{WFIRST} Microlensing in Optical Fields}
\label{sec:aux}}

Complementary ground-based survey observations of the \emph{WFIRST} microlensing fields have additional major benefits. First, ground-based surveys will be sensitive to planets in the wings of the events, in particular during the roughly 180 days (or 110 days) of a given year when \emph{WFIRST} is not observing. Of course, the ground-based sensitivity to planets will be much lower than would be the case for \emph{WFIRST} if it were observing. However, since it is not, this opens up completely new parameter space. While in some sense this parameter space is already available from ground-based surveys alone, the \emph{WFIRST} sample is unique in its sensitivity to low mass planets. Hence, the addition of a ground-based survey would significantly enhance our understanding of the relation between the relatively close-in terrestrial planets that \emph{WFIRST} is sensitive to and the more distant ice and gas giants that ground-based surveys are most sensitive to.

Second, it is likely that the characterization of microlensing sources and lenses (as well as other field stars) will benefit from follow-up observations, either individually or on a systematic basis. Such characterization is likely to benefit from the option of utilizing optical bands, and this would be greatly facilitated by observing low-to-moderate extinction fields.

\section{Discussion} \label{sec:discussion}

We propose here to augment the \emph{WFIRST} microlensing campaigns with simultaneous observations from a ground-based network of wide-field survey telescopes, in order to 1) enable 1-D microlens parallax measurements over the entire mass range $M\gtrsim M_\oplus$, and 2) yield 2-D parallax measurements for a significant fraction of short-timescale $(t_\e\lesssim3$~days) events and planetary/binary events. The 1-D parallax measurements can be used to produce complete solutions (mass, distance, transverse velocity) of the luminous lens systems ($M\gtrsim0.08M_\odot$), once combined with the measurements of the lens-source relative proper motion $\bmu_\rel$ that come from \emph{WFIRST} and/or ground-based follow-up high-resolution imaging. The 2-D parallax measurements will be especially useful for better understanding the substellar population. For the roughly 50\% of planetary and binary events with caustic crossings, and for a significant fraction of FFP events, the angular Einstein radius $\theta_\e$ is also measured, and thus these 2-D parallax measurements directly lead to complete solution of the lens (or the lens system). 

Our methodology applies as well to any other dedicated microlensing surveys that are conducted at L2, such as Euclid \citep{Penny:2013}.

Because \emph{WFIRST} launch is almost a decade in the future, one might in principle consider complementary observations from many different ground-based networks and/or individual telescopes.  For example, one might consider a network of infrared and/or optical telescopes of various apertures and field sizes \citep[e.g., similar to the ground-based network to support the \emph{K2} microlensing campaign,][]{Henderson:2016}. Here, we have initiated such an investigation by considering an optical, KMTNet-like network, simply because these telescopes exist and it would not be trivial to build a new network with substantially greater capability. The KMTNet microlensing campaign is scheduled to last 5 years and thus is expected to end before the launch of \emph{WFIRST}. Hence, coordination of KMTNet in particular with \emph{WFIRST} would also require advance planning, and this is another reason to begin serious evaluation of the benefits of such complementary observations now.

At first sight, it might appear that KMTNet would be ill-matched to \emph{WFIRST}. From the standpoint of finding planets, \emph{WFIRST} will be vastly superior.  However, the problem of augmenting \emph{WFIRST} observations in order to measure parallaxes given a $\sim$0.01 AU displacement from \emph{WFIRST} is far less challenging than finding planets from the ground.  Hence, as we have shown (see, e.g., Figures~\ref{fig:fullpar} and \ref{fig:piepar}), even 1.6m wide-field telescopes can yield excellent results for a very large fraction of events for which \emph{WFIRST} is sensitive to planets, including FFPs. The \emph{WFIRST} saturation magnitude might be another potential concern when complementing \emph{WFIRST} with ground-based 1.6m telescopes. However, \citet{Gould:2015} have shown this is not the case for near infrared cameras (see their Figure~1).


\acknowledgments

We would like to thank Scott Gaudi, Dave Bennett, Chung-Uk Lee, and Matthew Penny for discussions and comments. We also thank the Max Planck Institute for Astronomy for its hospitality. WZ and AG acknowledge support from NSF grant AST-1516842.

\appendix
\section{Fisher Matrix Analysis for General Cases} \label{sec:app1}

The general point-lens point-source light curve is described by Equation~(\ref{eqn:plps}). With this as well as Equations~(\ref{eqn:amp}), (\ref{eqn:bij_general}), (\ref{eqn:partials_general}), and (\ref{eqn:dadu_general}), the Fisher matrix can be evaluated
\begin{equation} \label{eqn:fisher}
b_{ij} = \frac{\Gamma }{\sigma_0^2} \left( \begin{array}{cccc}
        t_\e^{-1}F_\s^2 C_0 & 0 & 0 & 0 \cr
        0 & t_\e u_0^2 F_\s^2 C_1 & -u_0 F_\s^2 C_0 & t_\e u_0 F_\s C_2 \cr
        0 & -u_0 F_\s^2 C_0 & t_\e^{-1} F_\s^2 C_3 & -F_\s C_4 \cr
        0 & t_\e u_0 F_\s C_2 & -F_\s C_4 & t_\e C_5 \cr
\end{array} \right)\ ,
\end{equation}
where
\begin{equation}
\left\{ \begin{array}{rcl}
        C_0 &\equiv& \int A'^2 \tau^2 u^{-2} d\tau \cr 
        C_1 &\equiv& \int A'^2 u^{-2} d\tau \cr 
        C_2 &\equiv& \int A'(A-1)u^{-1} d\tau \cr 
        C_3 &\equiv& \int A'^2 \tau^4 u^{-2} d\tau \cr 
        C_4 &\equiv& \int A'(A-1) \tau^2 u^{-1} d\tau \cr 
        C_5 &\equiv& \int (A-1)^2 d\tau 
    \end{array}\right.\ .
\end{equation}
Then the covariance matrix is
\begin{equation}
c_{ij} = b_{ij}^{-1} = \frac{\sigma_0^2 t_\e}{\Gamma F_\s^2} \left(\begin{array}{cc}
        C_0^{-1} & 0 \cr
        0 & \mathcal{A}
\end{array}\right)\ ,
\end{equation}
where
\begin{equation}
    \mathcal{A} \equiv \frac{1}{D} \left(\begin{array}{ccc}
            \frac{C_3C_5-C_4^2}{t_\e^2u_0^2} & \frac{C_0C_5-C_2C_4}{t_\e u_0} & \frac{F_\s(C_0C_4-C_2C_3)}{t_\e^2 u_0} \cr
            \frac{C_0C_5-C_2C_4}{t_\e u_0} & C_1C_5-C_2^2 & \frac{F_\s(C_1C_4-C_0C_2)}{t_\e} \cr
            \frac{F_\s(C_0C_4-C_2C_3)}{t_\e^2u_0} & \frac{F_\s(C_1C_4-C_0C_2)}{t_\e} & \frac{F_\s^2(C_1C_3-C_0^2)}{t_\e^2}
\end{array} \right)\ ,
\end{equation}
and
\begin{equation}
    D \equiv 2C_0C_2C_4+C_1C_3C_5-C_0^2C_5-C_1C_4^2-C_2^2C_3\ .
\end{equation}
Therefore,
\begin{equation} \label{eqn:sigmas}
    \left\{ \begin{array}{rcl}
    \sigma(t_0) &=& \sqrt{\frac{t_\e}{\Gamma}} \frac{\sigma_0}{F_\s} C_0^{-1/2} \cr
\sigma(u_0) &=& \frac{1}{\sqrt{\Gamma t_\e}} \frac{\sigma_0}{F_\s u_0} \sqrt{\frac{C_3C_5-C_4^2}{D}} \cr
    \sigma(t_\e) &=& \sqrt{\frac{t_\e}{\Gamma}} \frac{\sigma_0}{F_\s} \sqrt{\frac{C_1C_5-C_2^2}{D}} \cr
    r(u_0,t_\e) &=& \frac{C_0C_5-C_2C_4}{\sqrt{(C_3C_5-C_4^2)(C_1C_5-C_2^2)}}
    \end{array}\right.\ .
\end{equation}
These expressions apply to both ground-based observations and \wfirst~observations (below sky limit). Here $r(u_0,t_\e)$ is the correlation coefficient between $u_0$ and $t_\e$, and only depends on $u_0$. We numerically find that $-1\le r(u_0,t_\e) \le -0.977$ for $u_0 \ge 0$ with the maximum achieved at $u_0=0.127$, and that $r(u_0,t_\e)\rightarrow0.997$ as $u_0\rightarrow1.0$. Therefore, parameters $u_0$ and $t_\e$ are strongly anti-correlated.

We are interested in $\sigma(t_0)$ and $\sigma(t_\eff)$, which are written in forms of Equations~(\ref{eqn:sigt0}) and (\ref{eqn:sigteff}). The former is already given in Equation~(\ref{eqn:sigmas}), and we further write it in a closed form
\begin{equation} \label{eqn:sigt0_app}
\sigma(t_0) = \sqrt{\frac{8}{\pi}\frac{t_\e}{\Gamma}} \frac{\sigma_0}{F_\s} u_0^{3/2} g(u_0)\ ,
\end{equation}
where
\begin{eqnarray} \label{eqn:gu0}
    g(u_0) &\equiv& \sqrt{\frac{\pi}{8}} u_0^{-3/2} C_0^{-1/2} \cr
    &=& \left[1-3u_0^2-3u_0^4+\frac{35+21u_0^2+3u_0^4}{(1+4/u_0^2)^{3/2}} \right]^{-1/2}\ .
\end{eqnarray}
There are two ways to determine $\sigma(t_\eff)$: by ground-based observations alone, and by incorporating the \wfirst~constraint on $t_\e$. We discuss these two cases in turn.

The ground-based observations alone yield
\begin{eqnarray} \label{eqn:sigteff1}
    \sigma_1 (t_\eff) &=& \sqrt{ t_\e^2 \sigma^2(u_0) + u_0^2 \sigma^2(t_\e) + 2t_\e u_0 r \sigma(u_0) \sigma(t_\e)} \cr
    &=& \sqrt{\frac{8}{\pi}\frac{t_\e}{\Gamma}} \frac{\sigma_0}{F_\s} u_0^{3/2} h_1(u_0)\ .
\end{eqnarray}
Here $r=r(u_0,t_\e)$, and
\begin{equation} \label{eqn:h1u0}
    \begin{array}{l}
    h_1(u_0) \equiv 
\sqrt{
{\pi\over8} \frac{(C_1C_5-C_2^2)u_0^4+2(C_0C_5-C_2C_4)u_0^2+(C_3C_5-C_4^2)}{u_0^5D} 
}.
\end{array}
\end{equation}
Note that one can show analytically that $h_1(0) = 1$.

To incorporate the constraint on $t_\e$ from \wfirst, we initially assume that $t_\e$ is known perfectly from \wfirst. This assumption has two requirements that must be specifically evaluated: 1) $\sigma(t_\e)$ from \wfirst~is extremely small compared to $\sigma(t_\e)$ from the ground; 2) the difference between $t_\e$ as seen from Earth and \wfirst~is effectively small. We will quantify these further below.

With perfect knowledge on $t_\e$, $\sigma_2(t_\eff)=t_\e \sigma(u_0)$, one can derive from the general case Fisher matrix (Equation~(\ref{eqn:fisher}))
\begin{equation} \label{eqn:teff2}
    \sigma_2(t_\eff) = \sqrt{\frac{8}{\pi}\frac{t_\e}{\Gamma}}\frac{\sigma_0}{F_\s} u_0^{3/2} h_2(u_0)\ ,
\end{equation}
where
\begin{equation} \label{eqn:h2u0}
\quad h_2(u_0) = \left[ \frac{\pi}{8u_0^5} \frac{C_5}{C_1C_5-C_2^2} \right]^{1/2}\ .
\end{equation}
The functions $g(u_0)$, $h_1(u_0)$ and $h_2(u_0)$ are illustrated in Figure~\ref{fig:func_u0}.

We now investigate the condition for the assumption that $t_\e$ is perfectly known. The covariance matrix of $u_0$ and $t_\e$ from ground-based observations is
\begin{equation}
    \left(\begin{array}{cc}
            \sigma_\oplus^2(u_0) & r\sigma_\oplus(u_0)\sigma_\oplus(t_\e) \cr
            r\sigma_\oplus(u_0)\sigma_\oplus(t_\e) & \sigma_\oplus^2(t_\e)
    \end{array} \right)\ .
\end{equation}
Here $r$ is the correlation coefficient. Let $\sigma_\w(t_\e)$ be the constraint on $t_\e$ from \wfirst~observations. Then after some algebra one can derive the combined constraint on $u_0$ as
\begin{eqnarray} \label{eqn:sigu0_general}
    \sigma^2(u_0) &=& \frac{\sigma_\w^2(t_\e) + 2(1+r)\sigma_\oplus^2(t_\e)}{(1+r)\sigma_\w^2(t_\e)/2 + \sigma_\oplus^2(t_\e)} \sigma_\oplus^2(u_0) \cr
    &\approx& \left[\frac{\sigma_\w^2(t_\e)}{\sigma_\oplus^2(t_\e)}+2(r+1)\right] \sigma_\oplus^2(u_0) \cr
    &\rightarrow& 2(r+1)\sigma_\oplus^2(u_0)\ .
\end{eqnarray}
In the first approximation we have used the fact that $\sigma_\w(t_\e) \ll \sigma_\oplus(t_\e)$, and the last step further assumes a perfect knowledge of $t_\e$ (i.e., $\sigma_\w(t_\e)=0$). This assumption remains valid as long as
\begin{equation} \label{eqn:condition}
    \frac{\sigma_\w(t_\e)}{\sigma_\oplus(t_\e)} \ll \sqrt{2(r+1)}\ .
\end{equation}
A typical value for the right-hand side is $0.15$. Here $\sigma_\w(t_\e)$ can be not only the statistical uncertainty on $t_\e$ from \wfirst~observations, but also the systematic uncertainty on $t_\e$ as seen from Earth because of the relative velocity between \wfirst~and Earth. We discuss the latter case in the next section. In the former case, if the source as seen from \wfirst~is also below sky, then with $\sigma(t_\e)$ from Equation~(\ref{eqn:sigmas}) one easily finds that the above condition (Equation~(\ref{eqn:condition})) is always met. For example, for the typical (reddened) source colors ($I-H=1.8$) and our adopted normalizations\footnote{We adopt $\Gamma_\oplus=240~$day$^{-1}$ and $\Gamma_\w=100~$day$^{-1}$, and assume 1\% \wfirst~photometry at $H=20.2$ and 5\% ground-based photometry at $I=18$.}, as well as a sky noise limit of $H=22.3$
    we get $\sigma_\w(t_\e)/\sigma_\oplus(t_\e)
=(0.01/0.05)\times 10^{0.4(18-20.2-1.8)}\times 10^{0.2(20.2-22.3)}=
1/520\sim 0.002$. 
The situation is not so favorable if the source as seen by \wfirst~is above sky especially when it is toward the peak, but since the $t_\e$ measurement depends mostly on the wings of the light curve and all that is required is a factor $\ll$$0.15$, this condition is still likely to be satisfied in almost all cases.

\section{Difference between $t_{\e,\oplus}$ and $t_{\e,\w}$} \label{sec:app2}

The difference between timescales as observed from Earth and \wfirst~is the systematic uncertainty in imposing \wfirst~$t_\e$ information on ground-based observations. Below we show that this difference is negligible for the majority of \wfirst~events that can be observed by ground-based surveys.

We first denote $\bdv{v}_{\oplus,\perp}$ and $\bdv{v}_{\w,\perp}$ as the velocities of Earth and \wfirst~transverse to the line of sight, and $\Delta \bdv{v}_\perp \equiv \bdv{v}_{\w,\perp}-\bdv{v}_{\oplus,\perp}$. Then
\begin{equation} \label{eqn:tediff}
    \frac{\Delta t_\e}{t_\e^2} \approx \left|\frac{1}{t_{\e,\oplus}}-\frac{1}{t_{\e,\w}}\right| = \left| \frac{|\bdv{\tilde{v}}_{\rm geo}|}{\tilde{r}_\e} - \frac{|\bdv{\tilde{v}}_{\rm geo}+\Delta \bdv{v}_\perp|}{\tilde{r}_\e} \right| \ ,
\end{equation}
where $\tilde r_\e= \au/\pi_\e$ is the projected Einstein radius and
$\bdv{\tilde{v}}_{\rm geo}$ is the transverse velocity of the event in the geocentric frame. By the triangle inequality $|\bdv{\tilde{v}}_{\rm geo} + \Delta \bdv{v}_\perp| \le \tilde{v}_{\rm geo}+\Delta v_\perp$, we can rewrite the above equation as
\begin{equation} \label{eqn:tediff2}
    \frac{\Delta t_\e}{t_\e^2} \lesssim \frac{\Delta v_\perp}{\tilde{r}_\e} = \frac{\Delta v_\perp}{t_\e\tilde{v}} \ .
\end{equation}
The orbit of \wfirst~is very likely to be a Lissajous orbit around L2. For our purpose, we simplify it as a circular orbit with period $P_\sat=180$~days and radius $R_\sat=2.7\times10^5$ km around L2. Thus, we find
\begin{equation}
    \Delta v_\perp \le \frac{2\pi R_\sat}{P_\sat}+\Omega_\oplus D_\sat |\sin\psi| \le 0.29~\kms\ .
\end{equation}
where we have taken into account that $|\psi|\le35^\circ$ during each campaign. The difference in $t_\e$ is then limited to
\begin{equation}
    \frac{\Delta t_\e}{t_\e} \lesssim \frac{\Delta v_\perp}{\tilde{v}} < 2.9\times10^{-4} \left(\frac{\tilde{v}}{10^3~\kms}\right)^{-1}\ .
\end{equation}
This means that even when \wfirst~can constrain its own timescale $t_{\e,\w}$ extremely well, the timescale of the same event as seen from Earth is still uncertain within up to $0.024\%$ ($0.086\%$) for typical Bulge (disk) events.

However, as long as the constraint on $t_{\e,\oplus}$ from ground-based observations alone, $\sigma_{\rm g}(t_\e)$, is significantly worse than the systematic uncertainty $\Delta t_\e$ derived above, $t_{\e,\oplus}$ can be treated at ``perfectly'' known and then Equation~(\ref{eqn:teff2}) applies. Quantitatively, this requires $\sigma_{\rm g}(t_\e) \gg \Delta t_\e /\sqrt{2[1+r(u_0,t_\e)]}$, or
\begin{eqnarray}
    \frac{1}{\sqrt{\Gamma_\oplus t_\e}} \frac{\sigma_{0,\oplus}}{F_{\s,\oplus}} \sqrt{\frac{C_1C_5-C_2^2}{D}} &\gg& \cr 1.5\times 10^{-3} \left(\frac{\tilde{v}}{10^3~\kms}\right)^{-1} 
    &\times& \left(\frac{1+r(u_0,t_\e)}{0.02}\right)^{-1/2}\ .
\end{eqnarray}
Numerically we find that
\begin{equation}
    \sqrt{\frac{C_1C_5-C_2^2}{D}} \ge 30 u_0^{3/2}
\end{equation}
for $u_0>0$ and that the minimum is achieved at $u_0=0.43$, based on which we can further write the above inequality as
\begin{eqnarray}
    \frac{\sigma_{0,\oplus}}{F_{\s,\oplus}} &\gg& 7.5\times10^{-4} \left(\frac{t_\e}{1~\rm day}\right)^{1/2} \left(\frac{\Gamma_\oplus}{240~\rm day^{-1}}\right)^{1/2} \cr
    &\times& \left(\frac{\tilde{v}}{10^3~\kms}\right)^{-1} \left[\frac{u_0^3(1+r(u_0,t_\e))}{0.02}\right]^{-1/2}\ .
\end{eqnarray}
This is the condition of Equation~(\ref{eqn:condition}) specified in the systematic limit regime.

For $u_0\ge0.2$, the last term in the above inequality introduces a factor $\le10$, and thus for \wfirst~targets observed with ground-based telescopes, this condition can almost always be satisfied. For smaller $u_0$, the above condition may not be satisfied for relatively bright sources, but the breaking down of the assumption does not make much difference, as the case of ``unknown $t_\e$'' ($h_1(u_0)$ curve in Figure~\ref{fig:func_u0}) is only worse by a factor $\le1.7$ compared the case of ``known $t_\e$'' ($h_2(u_0)$ curve in Figure~\ref{fig:func_u0}).


\end{document}